\documentclass[sigconf]{acmart}

\usepackage{booktabs} 

\usepackage{graphicx}
\usepackage{listings,xcolor}

\usepackage[caption=false]{subfig}
\usepackage{soul}
\usepackage{hyperref}

\renewcommand\footnotetextcopyrightpermission[1]{} 
\begin{document}
\title{Complex Event Processing of Health Data in Real-time to Predict Heart Failure Risk and Stress}


\author{Sandeep Singh Sandha}
\affiliation{%
  \institution{Department of Computer Science, UCLA}
}

\author{Mohammad Kachuee}
\affiliation{%
  \institution{Department of Computer Science, UCLA}
}

\author{Sajad Darabi}
\affiliation{%
  \institution{Department of Computer Science, UCLA}
}

\begin{abstract}
In this paper, we develop a scalable system which can do real-time analytics for different health applications. The occurrence of different health conditions can be regarded as the complex events and thus this concept can be extended to other use cases easily.
Large number of users should be able to send the data in real-time, and should be able to receive the feedback and result. 
Keeping the requirements in mind we used Kafka and Spark to develop our system. In this setting, multiple users are running Kafka producer clients, which are sending data in real-time. Spark streaming is used to process data from Kafka of different window sizes to analyze the health conditions. We have developed and tested the heart attack risk and stress prediction as our sample complex events detection use cases. We have simulated and tested our system with multiple health datasets.

\end{abstract}

\maketitle

\section{Introduction \& Background}
Real-time health data collection is very common these days. With the ubiquity of various low-cost health monitoring systems. This data is then acted on by various signal processing, and machine learning algorithms. The steps involved in extraction and inferences are similar across different applications. Researchers and engineers working with real-time signals perform similar pre-processing and processing steps prior to making inferences\footnote{https://md2k.org/}. The collected data can be used both in real-time and off-line to derive multiple inferences about the patients condition \cite{bates2014big}. However, applications in the healthcare domain are fairly limited due to the processing and network demands on the supporting infrastructure. A real-world healthcare application requires analyzing high-resolution sensor data in real time as well as data from other sources simultaneously, for many users at the same time. Processing the whole data on single machine locally is not practical due to computational limitations, reliability, scalability, failure/recovery and power consumption concerns. This is alleviated by using Kafka and Spark clusters, which are powerful distributed computing frameworks. 

Recently, there has been a great attention towards increasing the efficiency of the system through optimization of algorithms and the system implementation \cite{free2013effectiveness,kumar2013mobile,steinhubl2013can}. However, these methods only suggest solutions for specific problems. They involve design decisions that are difficult to generalize due to problem specific assumptions. To address this issue of application dependent implementations there is a need for processing platforms that are efficient enough to operate under real-world hardware and software constraints. Also, at the same time to be general enough to support different problems and applications. This work suggests an architecture that tries to solve this problem using intelligent distribution of the computational load using a publication/subscription scheme.

In this regard, the project's goal is to design and test an architecture which can scale to handle very large number of users and can act as a platform for processing real-time health analytics/inferences. We have developed our system using apache Kafka \footnote{https://kafka.apache.org/} and apache Spark \footnote{https://spark.apache.org/}. Both Kafka and Spark run on the cluster and have the capability to scale easily. Further, both frameworks are robust to multiple system failures. To test our system we have simulated the health data streams by means of Kafka producers and have implemented real-time heart risk and stress prediction use cases. The project uses the pub-sub systems and cluster computing framework to derive the complex event processing on top of health data.

The rest of the report is organized as follows. First we introduce the system design and architecture details. In this section we discuss different components of the system. In next section we explain the two use cases in detail. Next we discuss our prototype experiments and results. Last we discuss the future work and conclude the project.
\begin{figure*} 
\includegraphics[width=0.80\textwidth]{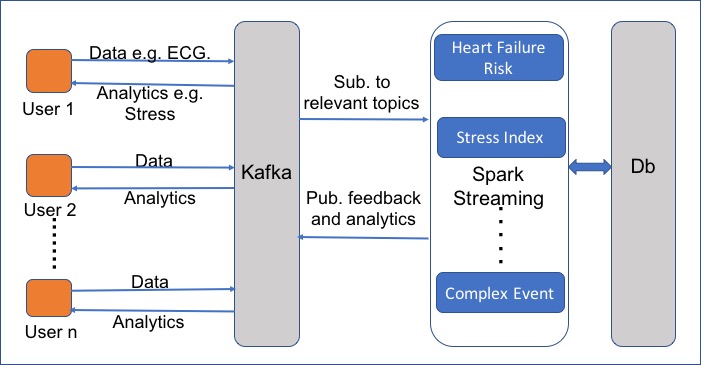} 
\caption{The system architecture\label{fig:SystemMain}} 
\end{figure*}

\section{System Design}
The system is designed using Kafka and Spark. Spark streaming is used to receive the data from Kafka topics. Spark streaming\footnote{https://spark.apache.org/streaming/} is also used to run the heart failure risk and stress index prediction over the windows of data streams.

\subsection{Apache Kafka}
Apache Kafka is a streaming platform which enables users to publish data and also subscribe to different streams of records. Kafka stores the streams in a fault tolerant way. It is used to build reliable real-time streaming data pipelines. Further, it runs as a cluster and implements the concept of topics or feed name which records are published to. In our current system abstraction, every user is publishing its data to a particular topic designated for the stream. Each topic internally maintains all the records for the configurable retention period. This capability can be used for health conditions that require using past data. 

\subsection{Apache Spark}
Apache spark is a big data processing framework that runs on clusters. It provides the promise of better reliability and fault tolerance along with in memory processing features. The different restricted transformations and queries are supported on the datasets, which are inherently parallel and fault tolerant. Spark supports the streaming library to build and scale streaming applications. It enables data injection from multiple sources including Kafka. It also provides an abstraction of DStreams which contains batches of data created from the real-time streaming source. The DStream is represented as a sequence of resilient distributed dataset also called RDDs, then the basic spark operations can be performed on these constructs.

\subsection{System Architecture}
The system architecture is shown in the figure \ref{fig:SystemMain}. Different users act as data producers and consumers in the Kafka model. Users publishes their health data for further processing. In this system the assumption is that users are aware of the different topics available for different types of sensory data. For example ECG data might be published to a topic dedicated to it, and BP data to a different topic and so-on. Kafka takes care of the scalability of the system as the number of users may grow or shrink. It also serves as a data retention, storage and forwarding interface.

Different types of complex events defining the health conditions are running on the Spark cluster. For example in figure \ref{fig:SystemMain} we have shown two such events of heart failure risk and stress index detection. These events are running as separate jobs using Spark streaming on the Spark cluster. Thus this system is extensible where multiple such jobs can be created or aborted depending on the desired analytics to be processed. Since, these jobs are running on the spark cluster they are scalable and are fault tolerant. The complex events considered here are explained in more detail in the use cases section.
	Each of the jobs are internally subscribed to the relevant health data which is done by means of Spark streaming. DStream are created from the multiple Kafka topics and are processed in each of these. The real-time generated output can be saved to the database and can also be communicated back to the user via Kafka. For sending results back to users, the jobs publishes back to the Kafka. Different users are also subscribed to the Kafka to listen to the relevant topics for the analytics results.

\begin{figure} 
\includegraphics[width=0.5\textwidth]{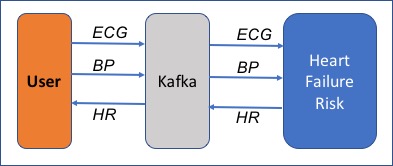} 
\caption{Simplified data pipeline of Heart Risk prediction\label{fig:HR_DP}} 
\end{figure}

\subsection{Data Pipeline}
The simplified data pipeline for the heart risk failure is shown in figure \ref{fig:HR_DP}. The algorithms used in the prediction for each use case are discussed in more detail in the next section. The user publishes the ECG and BP data to the Kafka. Heart Risk predictor is running as spark job, and uses spark streaming to continuously listen the Kafka topic for the user data. It is subscribed to two topics, one for ECG and one for BP. The Heart Risk Predictor publishes its output to the Kafka topic to which the user is listening. The sample data format of the ECG is shown below where all the fields are self explainable.

\lstset{
    string=[s]{"}{"},
    stringstyle=\color{blue},
    comment=[l]{:},
    commentstyle=\color{black},
}

\begin{lstlisting}
{
  "UserID": "101",
  "DataType": "ECG",
  "ValueType": "DOUBLE",
  "Value": 82.28,
  "TimeStamp": 1498004502
}
\end{lstlisting}

\section{Use Cases}
In this section we introduce two different use cases of the proposed system which are heart failure risk prediction and stress measurement from vital signals. In order to simulate the real-world complex event processing applied to each use case, recorded Electrocardiograph (ECG) and Blood Pressure (BP) signals from the Physionet online dataset \cite{goldberger2000physiobank} were used as a data source. Specifically, we have created two different Kafka producers corresponding to each signal (i.e., ECG and BP) that publishes time-stamp and signal amplitude, key value pairs at a constant rate equal to the signal sampling rate which is 500 Hz.

\subsection{Heart Failure Risk Prediction}
Heart failure (HF) or congestive heart failure (CHF) occurs when the heart is not capable of pumping enough blood in the cardiovascular system that is crucial for other body organs to work. CHF is one of the most common hospitalization reasons among people aged 65 years and older \cite{go2014executive}. Recently, there has been a great attention towards early detection and treatment of CHF using vital signals \cite{austin2013using}. However, little work has been done for real-time and online processing of these data sets. In this work, we present real-time CHF risk prediction as a potential use case of the introduced system.

On the data consumer side, we receive data from the ECG and BP producers in time windows of about 5 seconds. Prior to any analysis, signals of each window are preprocessed by statistically normalizing each window and applying appropriate Butterworth IIR filters.  Then, informative features were extracted from each window. Specifically, we have measured the minima and maxima values of each BP signal window and averaged them together to calculate the diastolic blood pressure (DBP) and systolic blood pressure (SBP) values. Extracting features from the ECG signal is slightly more involved. First, we normalize the ECG signal such that the minimum value is equal to zero and maximum value is equal to one. Then, the first derivative (i.e., difference) of the ECG signal is calculated and its zero crossing points are stored as potential local extrema points. Afterwards, ECG R-peaks inside the window are detected as the local maximums with the normalized amplitudes of more than 0.90 units. Finally, based on the relative occurrence position of other ECG key-points with respect to the ECG R-peak, we have detected P, Q, S, T time instances corresponding to each R-peak inside the window. It is noteworthy to mention that, as the signals we have used here are real-world vital signals, there are significantly corrupted or noisy signal parts that makes feature extraction extremely hard (if not impossible). In these cases, to reduce the systems complexity, we set the invalid feature values to a negative number to indicate the issue rather than using the common exception handing methods. It is the responsibility of processing blocks which use these features to check the validity of the extracted features before any further analysis.  

After the signal delineation phase, we calculate a number of high-level features which are shown to have a significant relationship with the occurrence of CHF including:
\begin{itemize}
\item Heat-rate value of more than 80 bpm.
\item QRS interval with more than 100 ms duration.
\item QRS interval with more than 120 ms duration.
\item QT interval with more than 410 ms duration.
\item Existence of the ST depression pattern.
\item Existence of the ST elevation pattern.
\item Existence of the inverted T-wave pattern.
\end{itemize}
Here, different time interval values are easily measured using the time instances of the occurrence of each ECG key-point inside the window as explained. Existence of different signal patterns such as ST depression and elevation patterns, and the inverted T-wave patterns were detected.  They are detected by comparing the amplitude of each ECG key-point with respect to each other and also the expected values for normal healthy ECG waveform. The exact analysis of each high-level feature and its contribution to CHF is beyond the scope of this work and interested readers are referred to related works in the literature.

At last, in order to predict the CHF risk values, we used a simple naive Bayes prediction algorithm to make the risk predictions. The model parameters were define based on the popularity of each high-level feature in CHF patients in the previous studies such as \cite{okin2006electrocardiographic} and \cite{vaclavik2014ecg}. Finally, we have normalized the predicted raw values using the maximum and minimum possible risk values such that the predicted values are representing the CHF risk percentages.


\subsection{Stress Index Measurement}
Stress is complex, as it is a subjective phenomenon and hence hard to define a precise measure. Stress is generally synonymous with distress, resulting from emotional, physical, and mental activities. It is the bodies method of responding to activities a person undertakes, and hence the stress response of every individual is different. This asks for methods capable of personalizing stress index to an individual person.

In the health literature there are various types of stress, acute and chronic stress. Acute (short term) stress results from demands of recent activities and near future. On the other hand, chronic (long term) stress results from pressure over long periods of times, which eventually lead to break down and abnormal behavior of a person. The type of stress and physiological cause of the person undergoing stress presents a challenge for processing. In this use case we predict short-term stress.

In the literature, there has been attempts on quantifying short term stress using heart rate variability (HRV), and frequency domain features of heart signals such as power in low frequency (LF) and power in high frequency (HF) \cite{GoldbergerH1273,ori1992heart,malliani1991cardiovascular}. It is noted that as the body undergoes stress the autonomic nervous system (ANS) has a direct role in the physical response of the body. The ANS is divided into the parasympathetic nervous system (PNS), and sympathetic nervous system (SNS). When the body undergoes stress due to physical activities, or mental load the SNS dominates the ANS. This results in lower HRV, and lower LF-HF ratio of the root mean square difference (RMMSD) of successive intervals extracted from the peaks of heart rate signals. In healthy conditions the PNS dominates the ANS and the respective values of HRV increases, and LF-LF ratio reduces. 
ECG signals are filtered and preprocessed as explained in the previous use case. 

The \textbf{features} to be extracted from each window are the peaks and HF, LF components. The R peaks are obtained using the peak detection algorithm as described previously. The indexes of the peaks are then used to calculate the HRV using RMSSD as a measure given by.
\begin{equation*}
RMSSD = \sqrt{\frac{\sum_1^NI[n] - I[n - 1]}{N - 1}}
\end{equation*}
where $I[n]$ denotes the index of the peak. In the literature it is suggested to use 1 minute windows for determining HRV values. 

The RR series are obtained from consecutive windows are used to obtain the LF and HF components. The PSD of the RR series is obtained using Fast Fourier Transform (FFT) method. The commonly used bands are very low frequency (VLF, 0 - 0.4 Hz), low frequency (LF,   0.4Hz - .15Hz) and high frequency (HF, 0.15 - 0.4). The LF component is obtained by integrating the frequency amplitudes between the range of 0.04 Hz - 0.15 Hz. Similarly, the HF component is obtained in the range of 0.15 Hz - 0.4 Hz \cite{Niskanen2004SoftwareFA}. The ratio $\frac{LF}{HF}$ provides an index for short-term stress. A high value is synonymous with high SNS activity, and a low value is associated with PNS dominant ANS. 

To predict the stress, a base-line measure of the person can be taken when he/she is sleeping. When the person is sleeping short term stressors are absent and hence a baseline can be obtained for the day. The baseline is then used to compare HRV values obtained through out the day. For distinguishing between physical stress, mental stress, and emotional stress it is required to take the context into account. The context can be obtained by using other sensors such as accelerometer, gyroscope, and GPS to determine the activities the person is undertaking. Given the context it is possible to better distinguish between good stressors and bad stressors.

\section{Use Cases Operations in Spark RDDs}
In this section we will discuss how to define the use cases in the spark map reduce framework and the other transformations and operations supported by it. The goals behind these definitions is that, these operations are highly parallel and fault tolerable. Also in future these operations may be individually moved to different places for the reason of faster processing, reliability and security. 

Kafka producer is sending the data in the Json array format using the windows of sizes required for the heart risk measurement and stress index measurement. We created severals RDDs from the data received using the spark transformations. Logically each RDD is a different stages in the signal processing and are equivalent to the peaks in the ECG and BP signals, along with different operations on it.

\section{Experiments \& Results}
The prototype system was implemented using Java. The code of the implementation is available on the git\footnote{\url{https://github.com/sandeep-iitr/ComplexEventDetection_CS249}}. We simulated the multiple users by sending real-time data via different Java applications to the Kafka cluster. The implementation of both use cases is using Spark streaming in Java, and each is being run as a separate job on the local Spark cluster. The jobs do the real-time health analytics and produce the heart risk and stress index as output.

\begin{figure}[htp]
\subfloat[]{
  \includegraphics[clip,width=\columnwidth]{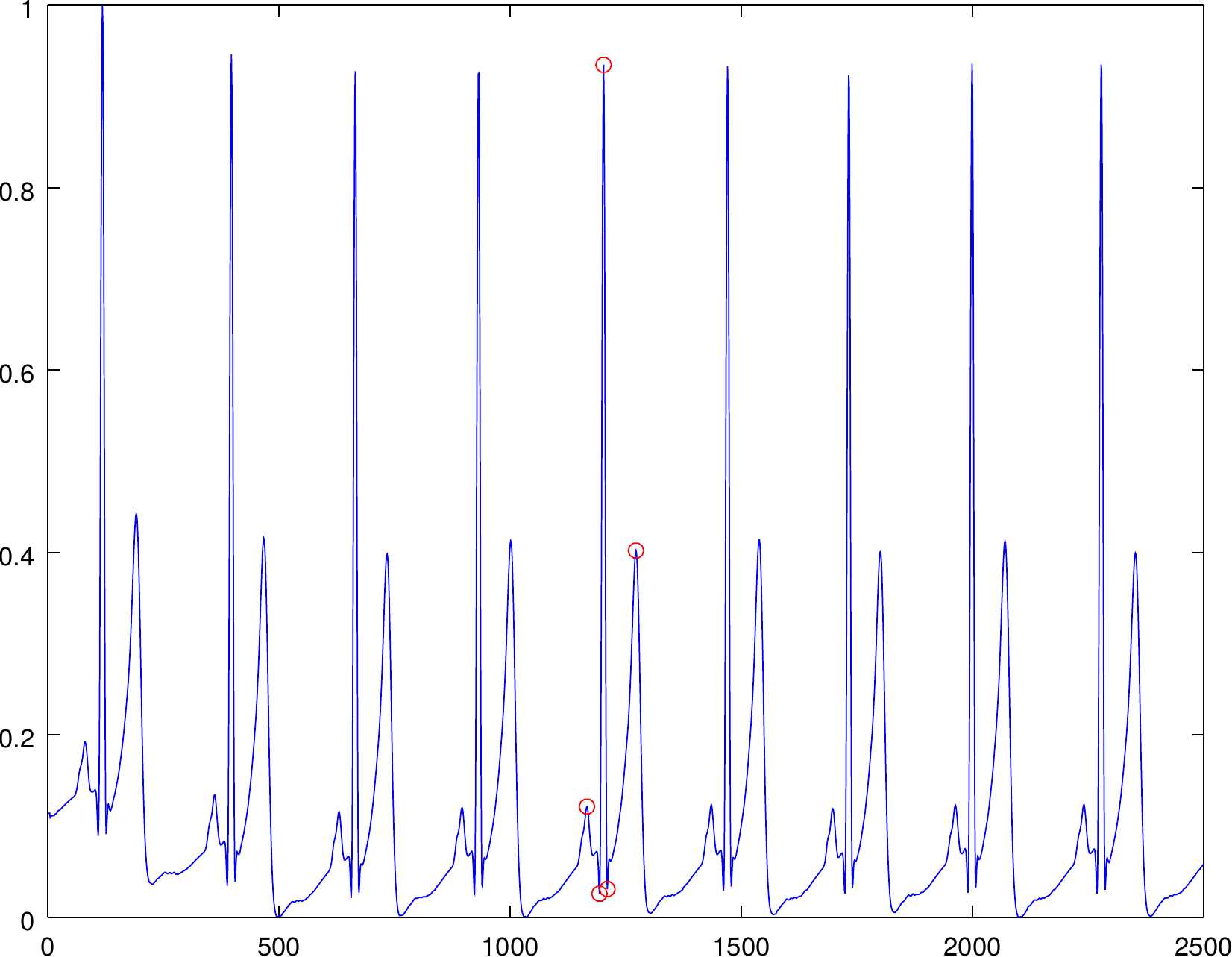}
  \label{fig:examples_1}
  }

\subfloat[]{
  \includegraphics[clip,width=\columnwidth]{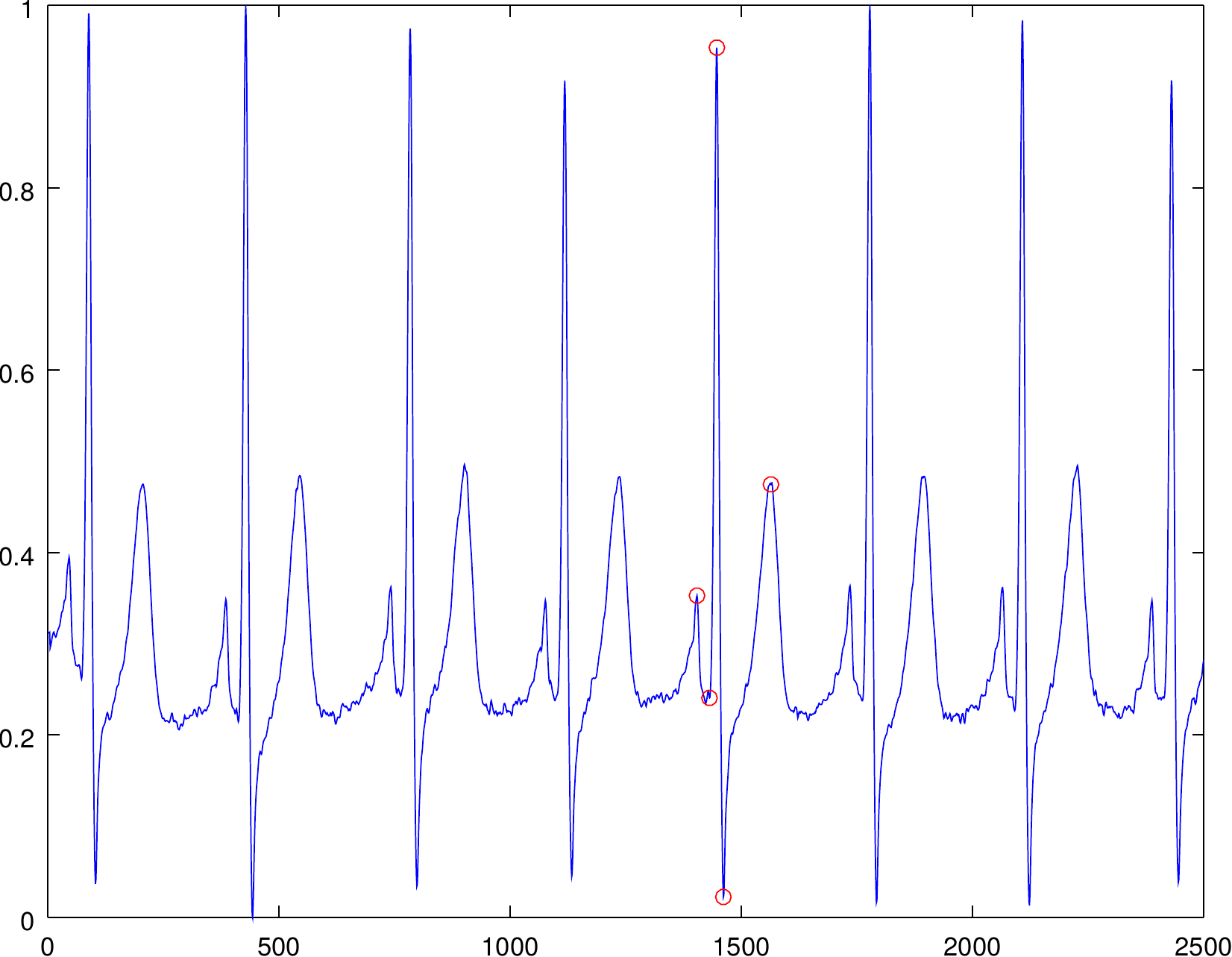}
  \label{fig:examples_2}
  }
  
\subfloat[]{
  \includegraphics[clip,width=\columnwidth]{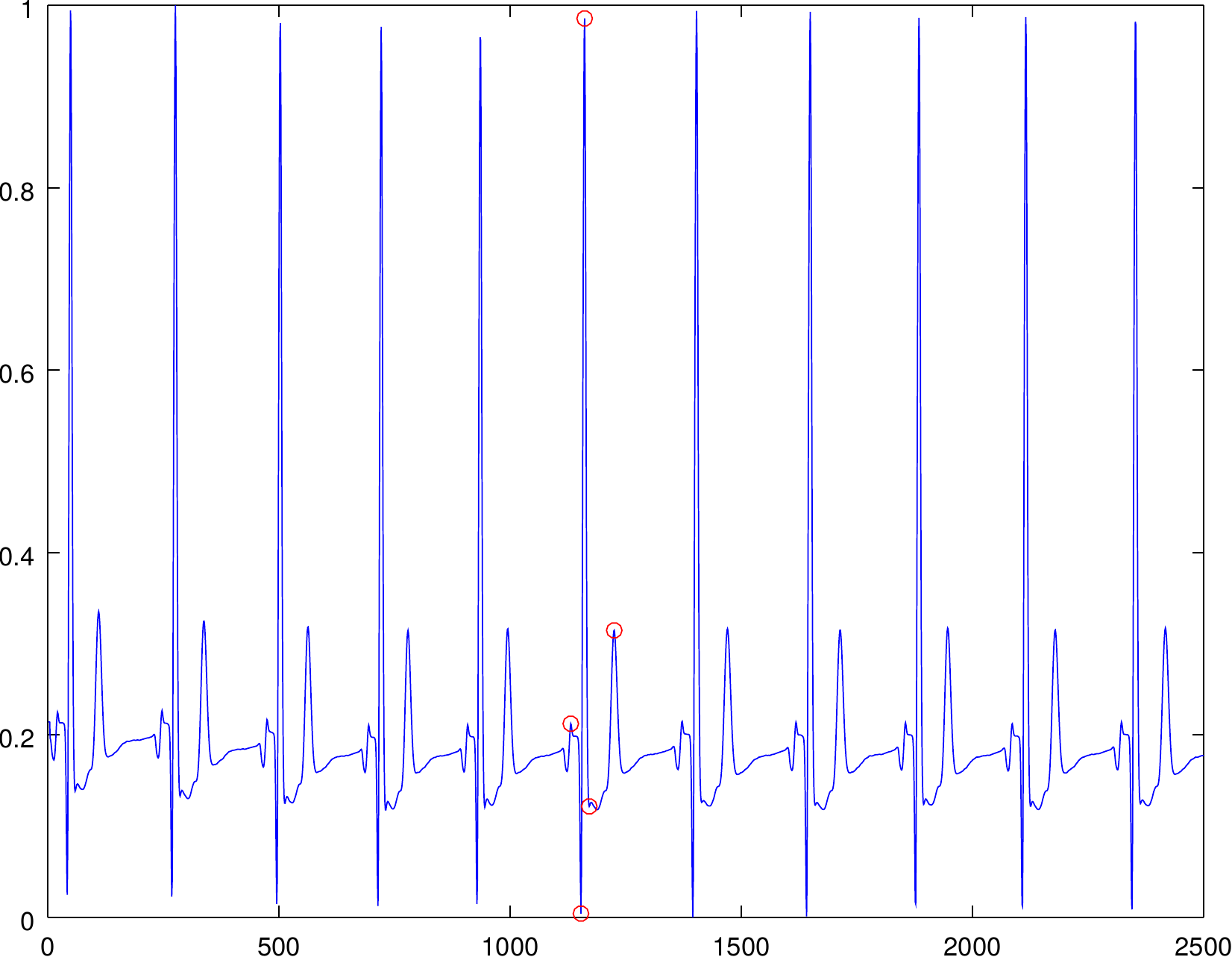}
  \label{fig:examples_3}
  }
\caption{Examples of three different ECG signals with different CHF risk values. (a) ECG signal with normal shape and predicted risk of $1.2\%$. (b) ECG signal with abnormal S and the predicted risk of $12.5\%$. (c) ECG signal with ST depression pattern and the predicted risk of $20.6\%$.}
\label{fig:examples}

\end{figure}

Figure \ref{fig:examples} shows three different ECG signal windows corresponding to different individuals. For each signal, the P, Q, R, S, and T ECG key-points are indicated with red circle markers inside each plot. Figure \ref{fig:examples_1} is an ECG signal with the waveform similar to a normal healthy person, and the predicted CHF risk by the proposed is about $1.2\%$. Figure \ref{fig:examples_2} is related to an ECG waveform with slightly abnormal S point, and the predicted risk in this case is about $12.5\%$. Finally, \ref{fig:examples_3} is a waveform with ST depression pattern and with the predicted risk of about $20.6\%$. As it is evident from this example, the proposed system is able to predict reasonable CHF risk values based on available data in real-time.

To calculate the stress index, the heart rate and HRV for each window is used. In this implementation the stress index corresponds to a change in HRV in consecutive windows. The consecutive difference of RR intervals are stored in a circular buffer before more windows are obtained. Once the circular buffer is full, at least 1 minute duration worth of RR interval differences is stored and is used to calculate the HRV for the corresponding window. If the HRV drops below the baseline, SNS is taking over the ANS and hence the stress index is increased. If the HRV increases PNS is more dominant and the stress index decreases. In this implementation the baseline is set to be the HRV obtained from the previous window. 

\begin{table}[h!]
\caption{Examples of the stress measurement use case}
\label{tab:stress}
\begin{center}
\begin{tabular}{ |c|c|c|c| }\hline
 Window \# & HR & HRV & Stress Index\\\hline
 1 & 138.5 & 107.53 & 0.1\\\hline
 2 & 143.3 & 105.00 & 0.2\\\hline
 3 & 147.3 & 104.59 & 0.3\\\hline
 4 & 148.73 & 98.86 & 0.4\\\hline
 5 & 148.3 & 97.02 & 0.5\\\hline
 6 & 143.3 & 98.47 & 0.4\\\hline
\end{tabular}
\end{center}
\end{table}

As depicted in Table \ref{tab:stress}, the stress index starts at 0.1, and increases to 0.5 before dropping back down. This is due to a decrease in the HRV value, whereas HR is relatively constant.

As the major focus of this work is to present a general architecture for the CEP implementation on mobile systems and use cases are used as examples that run on this system, detailed analysis and evaluation of the proposed use cases is out of the scope of this study.  



\section{Future Work}
In the prototype implementation we have implemented the two health use cases. These implementations prove that a general platform is capable of processing other applications of health inference and behavioral patterns. For instance, real-time prediction of smoking can be easily done using the proposed system architecture. We would also like to test this system on real scenarios, as in our current prototype implementation we simulated the user data using real datasets by running separate Java client applications.

More work can also be done to define the complex events in generalized way for the health applications, so that the different signal processing capabilities can be reused. A simple and expressive grammar can be proposed for the same which achieves the job of events semantic definitions, the required inputs, outputs and logic. However such a goal requires effort to understand the nature of different health analytics to develop a generalization.

\section{Conclusion}
In this project we worked on developing a framework to do health analytics in real-time.
By developing the framework using Kafka and Spark, we can handle large amounts of health data reliably and efficiently in a fault tolerable fashion. Individual components of system are designed to be highly scalable. Further, the system can be easily extended to support new use cases via running each as a separate spark job. We implemented a prototype system for running two use cases in real-time. Our prototype shows the capabilities of the proposed design. The design can be improved by adding the  data security and access control features so as to deploy with real users or to conduct studies.


\bibliographystyle{ACM-Reference-Format}

\end{document}